# Simulation of Physical Parameters for a Photoneutron Source

Xiaohe Wang[1,2,3], Longxiang Liu[1,2], Jifeng Hu[1,2], Jianlong Han[1,2], Pu Yang[1,2,3], Zuokang Lin[1;2], Guilin Zhang[1,2], Naxiu Wang[1;2;3], Xianzhou Cai[1,2,3], Hongwei Wang[1,2]*, and Jingen Chen[1,2,3]*

( 1. Shanghai Institute of Applied Physics, Chinese Academy of Sciences, Shanghai 201800, China;

2. CAS Innovative Academies in TMSR Energy System, Shanghai 201800, China;

3. University of Chinese Academy of Sciences, Beijing 100049, China )

**Abstract :** A compact photoneutron source (PNS), based on an electron linac was designed and constructed to provide required nuclear data for the design of Thorium Molten Salt Reactor (TMSR). Many local shielding are built to reduce the background of neutron and γ rays, making the location of the time of flight (TOF) detector be fixed at 6.2 m place. Under the existing layout, some physical parameters are very difficult to get by the experiments, while can be obtained by the Monte Carlo simulation method. However, for the deep penetration problem of the neutron and γ rays transport in the channel of PNS with complex geometry, the normal Monte Carlo method is inefficient since electron transport calculation need a large amount of computing time and neutrons have little contribution to the detector in far-source region. In this work, the subsection method is applied in the simulation for PNS, which divide the simulation process in two steps, recording the neutron and γ rays information passing through the source window in the first step and adopting the covariance reduction techniques in the second step. The simulated neutron flux and energy spectrum at the TOF detector place with the relative error 1.6% are well agreement with the experimental results, achieving an efficiency 23 times better than the normal method. This method is fast and efficient in predicting the physical parameters, providing a required verification and initiating the foreseen physics experiment.

PASC:

## 1. Introduction

The target accuracy of effective neutron multiplication factor uncertainty from nuclear data for the six generation IV nuclear energy systems is estimated to reach 0.3% which, of course, cannot be fulfilled by the existing nuclear data [1]. Many electron accelerator-driven neutron sources, such as GELINA [2], nELBE [3] and PNF [4], are built to improve the accuracy of nuclear data for the design of advanced nuclear power reactors. In order to meet the requirement of nuclear data for the Thorium Molten Salt Reactor (TMSR) nuclear system [5], a photoneutron source (PNS) driven by a 15 MeV electron linac has been built at Shanghai Institute of Applied Physics (SINAP) [6]. The main objective of this facility is to provide required neutron cross section measurements for some key nuclides concerning TMSR. Besides, benchmarking for evaluated nuclear data and validation for nuclear design code are expected to be performed at this facility.

The PNS is an electron accelerator-driven neutron source which consists of an electron accelerator, a photoneutron target and a detector system, all arranged in an experimental hall with space of 11m × 8 m, causing inevitably a considerably high background of neutron and γ rays in the hall. In order to reduce the influence of background on measurement, an elaborate shielding by considering various required source terms of background at different locations is constructed, making the detector system be fixed at 6.2 m

E-mail:wanghongwei@sinap.ac.cn

E-mail:chenjingen@sinap.ac.cn

place away from the photoneutron target, which restricts the physical parameters measurement of other place. Therefore, it is very important to find a method to get all parameters and confirm the performance of PNS as fast and effortlessly as possible. The Monte Carlo method can be used to simulate the parameters of this facility, but the normal method is practically impossible, because the computational expense is huge. In addition, the detector cell is too small relative to the geometry of PNS and located behind a significant amount of shields and moderators, leading the neutrons can be scored are very few. For other facilities, such as n_TOF at CERN [7], China Advanced Research Reactor [8] and so on, there are similar difficulties in the simulations. The subsection method, dividing the model into several parts and simulating in separate steps, is taken to get the physical parameters. This method can also be used in the simulation for PNS, but more actions are needed to improve the accuracy and efficiency of the simulation.

In order to determine the physical parameters and verify the system design of PNS in the commission period, the subsection method together with the variance reduction techniques based on Monte Carlo method (Sub-Var method) is performed to simulate the PNS by a two-step calculation and proved by comparing with the measurement results of neutron flux and energy spectrum at TOF detector place. At the same time, the detection efficiency of the $^6$LiF(ZnS) scintillator and some physical parameters along the flight path which can't be obtained by the experiments at present are simulated. This simulation method is helpful to provide guidance for the future experiments. The present work will give a description of TMSR-PNS in detail, followed by the experimental setup, measurements, simulation method and the results discussion.

## 2. Experimental setup and measurements

### 2.1 TMSR Photon Neutron Source

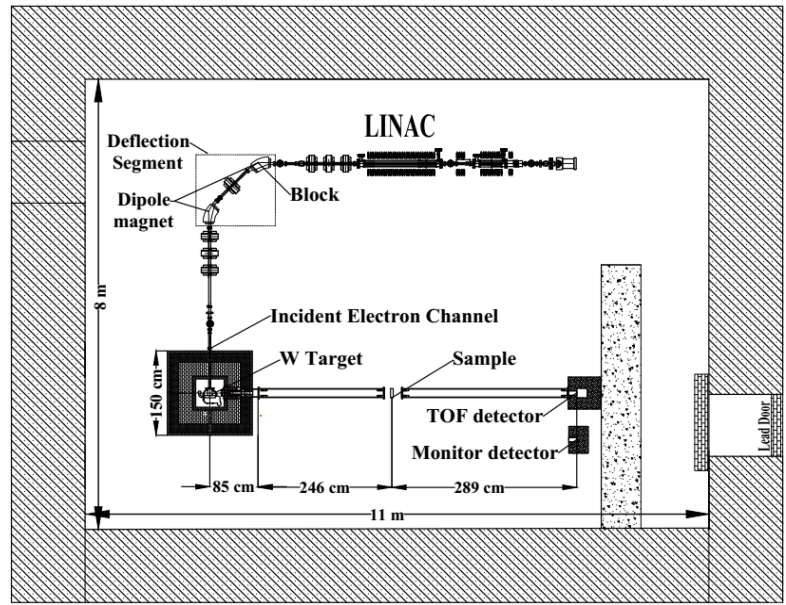

**Figure.1. Facility setup (vertical view) of TMSR-PNS**

The experimental arrangement of TMSR-PNS is shown in Fig.1. It is based on an electron linear accelerator (LINAC) producing electron beams with a typical beam operation mode characterized by 15 - 18 MeV average energy, 3 ns - 3 us pulse width, 10 - 260 Hz repetition rate (W) and 0.1 mA average current. The designed electron beam power on target is 1.5 kW. The electron beam hits the neutron target, the γ rays are produced in the bremsstrahlung process, then go on to produce neutrons by photoneutron reactions.

The photoneutron target is a cylindrical tungsten target with a diameter of 60 mm, a thickness of 48 mm and a purity of 98%. It is placed at the center of the target housing and welded on a water-cooled copper pedestal. In order to reduce the neutron and γ rays background, the photoneutron target is shielded by a box consisting of 5-cm aluminum, 25-cm lead, 15-cm polyethylene and 5-cm aluminum, in sequence. The neutron guide tube is aligned vertically to the electron beam while the γ rays guide tube is aligned with the electron beam forward direction which is blocked by lead plate at present. Neutrons from the target will go through a lead plate with a thickness of 5 cm for reducing the impact of γ rays, followed by a 10-cm thick polyethylene plate to thermalize the fast neutrons. The diameter of neutron beam is collimated from 10 cm to 5 cm by the Boron-loaded Polyethylene (PE-B, 5 wt.%) tube and the lead tube, as shown in Fig.2.

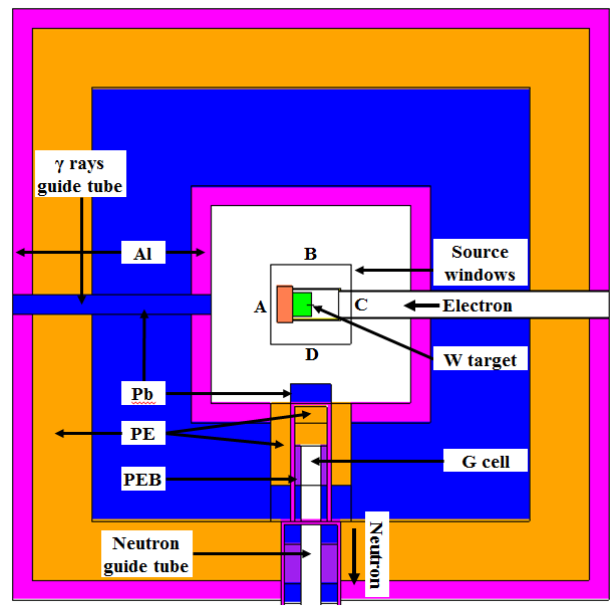

Figure.2. Simulating graph of target housing

After the neutrons come out from the guide tube, they fly to the detector through two stainless steel tubes with the same diameter of 14 cm and two different lengths, 246 cm and 289 cm. In the middle of two tubes, there is a 30 cm space reserved as the experimental target area where the sample changer is located. At both ends of each tube, two collimators, composed of lead and PE-B block with the same inner diameter and thickness of 5 cm, are installed to reduce the radius of the neutron beam to 5 cm. The total flight path is 6.2 m.

Two identical type neutron detectors are used in the experiments. One is TOF detector which is used to detect the transmitted neutrons from the tungsten target, located at the end of second neutron guide tube. Another is monitor detector which is placed at the lower left of TOF detector, and data from this monitor is used to monitor the neutron intensity during the experimental period.

**2.2 Neutron background**

The high neutron and γ rays background has serious affected the accuracy of the measurement results. In the commission period, the neutron background is mainly from two aspects. The first one is the electron beam loss at the accelerator deflection segment which is shown in Fig.1. When the electrons are accelerated to the design energy, the electron beam is deflected in $90^0$ by two dipole magnets due to the hall space constraint. Three beam blocks are used to sweep away those unwanted particles, while two of them are for removing low energy electrons and one is for removing high energy electrons. These lost electrons produced at the deflection segment will react with the surrounding materials and generate the neutrons and γ rays, part of them will enter the detector system. It is found that the loss ratio of the beam is about 30%. The second one is back forward neutrons relative to the incident electron direction from the tungsten target housing. The electron beam tube is directly connected to the target without any shield materials, the neutrons will fly to the hall through this tube.

In order to get the better shielding effect, it is necessary to calculate and analysis all kinds of shielding schemes. At this time, a fast and efficient simulation method is particularly important, that the Sub-Var method is applied in the shielding calculation. Finally, three local shields are taken to reduce the neutron and γ rays background. First, the accelerator deflection segment is shielded by 10-cm aluminum plate, 10-cm lead plate and 10-cm PE-B plate in sequence. Second, a L-type shielding wall which is made of concrete and PE-B with the same thickness (30 cm) and height (200 cm) is established to separate the detector system from the experimental hall. Third, the TOF detector is placed in a shielding box with $100 \times 100 \times 100$ cm$^3$ which is made of 30-cm thick PE-B. Neutrons can reach the detector through a channel with a diameter of 5 cm. The monitor detector is shielded by a box with 20-cm thick PE-B. The scheme of the shielding for this facility is shown in Fig.3. After the adoption of these shields, the neutron background is reduced to less than 2%, then the measurements of neutron cross section can be carried out.

**2.3 Neutron flux and energy spectrum measurement**

The neutron flux and energy spectrum at TOF detector place were measured in June 2016. The electron energy was 16 MeV and the error was limited to 2%. Three reaction targets were used for cycle testing and 300 s of measured time was taken for each sample. The first one was a set of notch filters contained Cd (purity 99.99%) with a thickness of 0.125 mm, Co (purity 99.9%) with a thickness of 0.05 mm, Ag (purity 99.95%) with a thickness of 0.1 mm, and In (purity 99.99%) with a thickness of 0.05 mm, which was used for energy calibration. The second one was a PE-B cube with a length of 12.8 cm which was used to block the neutrons from the target for the background measurement. The last one was a blank sample which is called open beam.

Because of high detection efficiency for thermal neutrons with low sensitivity to γ rays radiation, $^6$LiF(ZnS) scintillator which is made by Eljen company with the product code of EJ426HD2 was used in the experiments and mounted on an EMI9813 photomultiplier produced by ET enterprise Ltd. The scintillator has the form of a flat, white thin sheet consisting of a homogeneous matrix of fine particles of Lithium-6-Fluoride ($^6$LiF) and Zinc Sulfide phosphor (ZnS:Ag) compactly dispersed in a colorless binder. The scintillating matrix is 76 mm in diameter and 0.5 mm in thickness, wider than the neutron beam. The lithium is enriched in $^6$Li to a minimum of 95 at.%.

The TOF technique is used to measure the neutron flux at 6.2 m. The DAQ system which is used in the experiments had been described in detail in Ref. [9].The signal from the detectors is sent to the DAQ which is a CAEN DT5720 digitizer with four channels. The signal from TOF detector is connected to channel 1, the signal from beam intensity monitoring is connected to channel 2, and the RF signal from electron gun is connected to channel 3 as the start signal.

3. **Simulation method**

Simulating the geometry like PNS by normal Monte Carlo method is a very difficult and time-consuming task. For PNS, neutrons are produced by electron beam onto the tungsten target, transported through flight tubes and moderated or absorbed by light nuclei. These complex processes lead to more time to calculate. In addition, the solid angle is so small that very few neutrons can be scored by the TOF detector. The subsection method combined with the variance reduction techniques based on Monte Carlo method is taken to solve this problem. The simulation process is separated into two steps: i) construct a suitable window in the target housing as a new source for the next calculation, and record the information of particles transporting to this window through the direct calculation by SSW card; ii) combine with the variance reduction techniques and carry out the transport calculation by taking the window of first step as the new neutron source. The simulating scheme of PNS is shown in Fig.3.

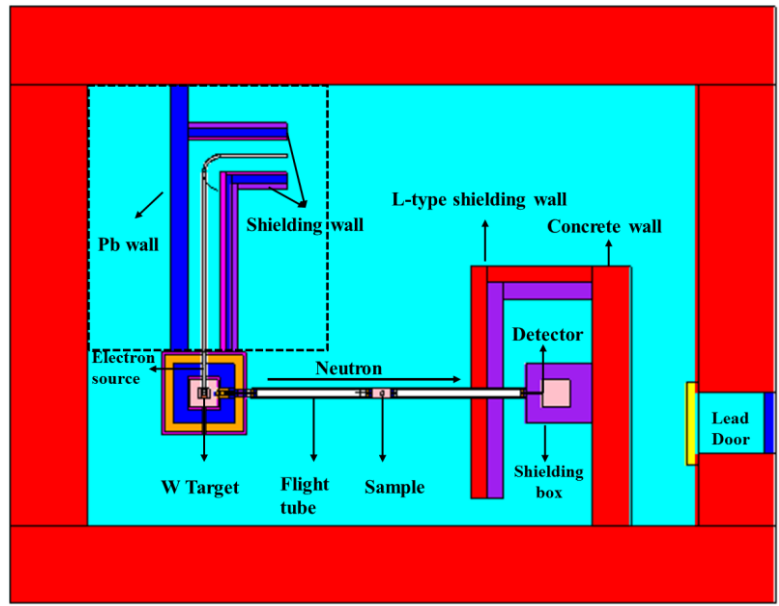

**Figure.3. Simulating graph of PNS**

In the first step, the source window is constructed at the center of target housing, which is composed of six surfaces. Four of six surfaces (A, B, C and D) are shown in the Fig.3, while two (E and F) are not shown. The advantage of this setup is that this source window can be used for the foreseen simulations with different conditions without changing the source window. Due to the application of local shields, the influence of neutron background on the detector which is produced by the accelerator deflection segment can be ignored. Therefore, an electron face source with a diameter of 5 mm is used to generate electrons, instead of the electron linac (in the dashed box, shown in Fig.3) in the simulation. The simulation code MCNP5 is used to complete a direct calculation with the detailed geometry of PNS. All information of neutrons and γ rays produced by 16 MeV electron beam onto the tungsten target coming out from the window are recorded.

In the second step, the previous neutron window is used as a new neutron source to continue the simulations without considering the electrons. All neutrons that reenter the source window are killed to prevent the neutrons coming out of the window to be counted again. Otherwise it will make the results larger. The variance reduction techniques are taken to increase the efficiency and accuracy of calculations [10], including geometry splitting and russian roulette, DXTRAN sphere and forced collision. DXTRAN sphere is one of the most useful techniques in the streaming problems. This technique can force the particles to a small region of space that they will be otherwise very unlikely to go. Considering the rarity and high weight of the particles that reach the DXTRAN sphere, which indicates their high importance to the tally result, it is advisable to increase the frequency of these events, and the forced collision method can be used to increase the frequency of collisions in specified cells.

In the simulations, the value of NPS which means how many particles to run is $2 \times 10^9$. The calculation is carried out in the TMSR Supercomputing Center. 16 processors are taken for the direction calculation, but only two processors are taken for the Sub-Var method due to the source window file is too large to carry out the parallel computing. Both of neutron data and photonuclear data are from ENDF/B-VII.0 [11]. The calculation temperature is room temperature. Considering the impact of thermal neutron scattering effect of the polyethylene, S (α，β) from ENDF/B-VII.0. is taken.

### 4. Results and Discussion

**4.1 Sub-Var method validation**

The measured data of neutron flux and energy spectrum in June 2016 are compared with the simulation results to verify the reliability of the Sub-Var method. The electron beam power on target is 1050W and the electron energy is 16 MeV.

In order to check the reliability of the source window setting, the neutron flux at G cell which is behind the PE moderator (shown in Fig.2) is calculated by the direct calculation and the Sub-Var method. The results is shown in Table.1 It is found that the difference of the calculated neutron flux between two methods is about 0.3%, which means the layout of the source window in the simulation is reasonable. The relative errors is 1.44 and 0.76, the figure of merit (FOM) which is used as the measure of calculation efficiency is 4.7 and 10, respectively. It is indicated that the calculation efficiency and accuracy of the Sub-Var method have been improved.

**Table 1 Simulation results of neutron flux at G cell with different methods**

|  | Neutron Flux (/s/cm$^2$) | Relative Error (%) | FOM |
|---|---|---|---|
| Direct Calculation | $1.311 \times 10^7$ | 1.44 | 4.7 |
| Sub-Var method | $1.315 \times 10^7$ | 0.76 | 10 |

The simulated neutron flux at TOF detector place by two methods is shown in Table2. It turns out that neutrons are hardly scored to the TOF detector by the direct calculation, but can be scored by the Sub-Var method. The simulated results pass all statistical checks [12], the relative error is 1.6% and the FOM is 4.3. The calculation time is reduced from 830 hours to 35 hours (the second step), increasing the calculation efficiency by 23 times. The simulated neutron flux is 85.0 /s/cm$^2$, which is 18.9% larger than the measured results (68.9 /s/cm$^2$, shown in Table 2).

**Table 2 Neutron flux at TOF detector at a power of 1050 W**

|  | Neutron flux (/s/cm$^2$) | Relative Error (%) | FOM | Time(h) |
|---|---|---|---|---|
| Direct Calculation | 0.0 | 0.0 | 0.0 | 830 |
| Sub-Var method | 85.0 | 1.6 | 4.3 | 35 |
| Measured result | 68.9 | 2.3 | -- | -- |

The simulated neutron energy spectrum which considering the detection efficiency is shown in Fig.4 together with the measured one. It is observed that the energy spectrum of PNS is in a wide energy domain, from 0.001 eV to 10 keV. In the low-energy region (below 0.006 eV), the simulated data shows a good agreement with the measured data. In the energy region of 0.006 eV to 0.02eV the simulated data is higher than the measured one obviously, but in the energy region of 0.02 eV to 10 eV it is slightly lower than the measured one. This is caused by the difference between the moderator used in the simulation and measurement. Above 10 eV, the deviation between the simulated data and the measured ones gradually increases. Due to the pulse-shape discrimination (PSD) method was applied for n/γ identification for TOF spectrum, there is a threshold for time of flight which will make the measured neutron counting lower than the actual one. In addition, in this energy region, the detection efficiency is quite low, less than 2%. Therefore, the discrepancy of neutron flux and spectrum between the simulation and measurement are acceptable.

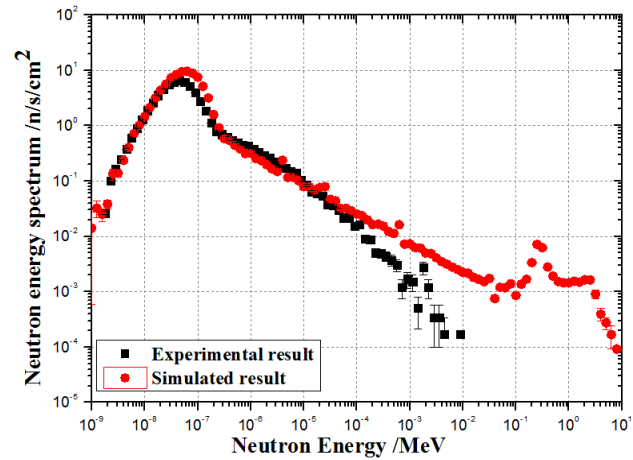

**Figure.4. Energy spectrum of simulation and measurement at 6.2 m in a power of 247 W**

The $^6$LiF(ZnS) scintillator has been calibrated in the Xi'an pulsed reactor (XAPR)[13] which is a typically thermal reactor. The calibration efficiency is about 57.35% and the uncertainty is about 10%. The detection efficiency of the $^6$LiF(ZnS) scintillator is also simulated by the Sub-Var method. The simulated detection efficiency in the same energy region is 60.9% which is larger than the calibrated value due to take into account the impact of the surrounding environment.

**4.2 Physical parameters calculation**

In the commission period, many local shields are taken to reduce the background of neutron and γ rays, leading the layout of the experimental hall be relatively fixed. If some physical parameters are needed to verify the design of PNS and provide more information for the future the experiments through measurements, it is necessary to reconstruct the device and re-shield, which is very costly and time-consuming. Therefore, the Sub-Var method which has been proved can be used to obtain these parameters.

The neutron flux at the target housing (30 cm from the target) and the end of guide tube (85 cm from the target), where can be used to perform some experiments, are simulated. The result is shown in Fig.5. It is found that, as the distance between the detector and target increases, the neutron flux is decreased rapidly. When the power reaches 1.0 kW, the neutron flux at the target housing is about $10^8$ n/s/cm$^2$ that the irradiation tests can be performed, and at the end of neutron guide tube is about $10^4$ n/s/cm$^2$ that the neutronics integral experiment which needs higher neutron flux can be carried out with the proper transformation.

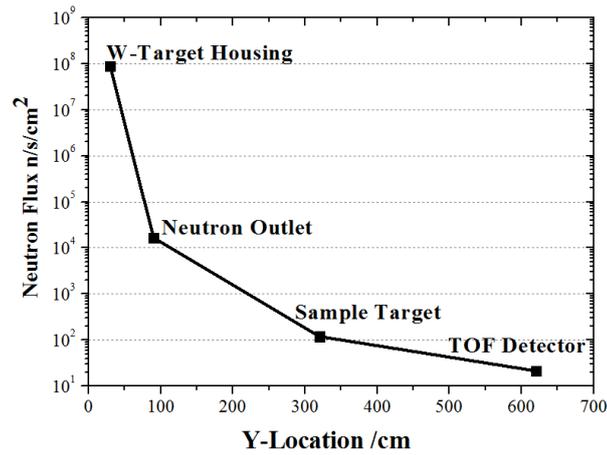

**Figure.5. Neutron flux at different place away from photoneutron target**

When the electron energy is 16 MeV, the neutron yield of tungsten target is $2.7 \times 10^{11}$ n/s/kW. It is agreed well with the results ($3 \times 10^{11}$ n/s/kW ) performed by IAEA [14]. Fig.6 shows that the neutron energy spectrum of target is a typical fast energy spectrum, and the energy peak of neutrons produced by electrons is at about 1 MeV.

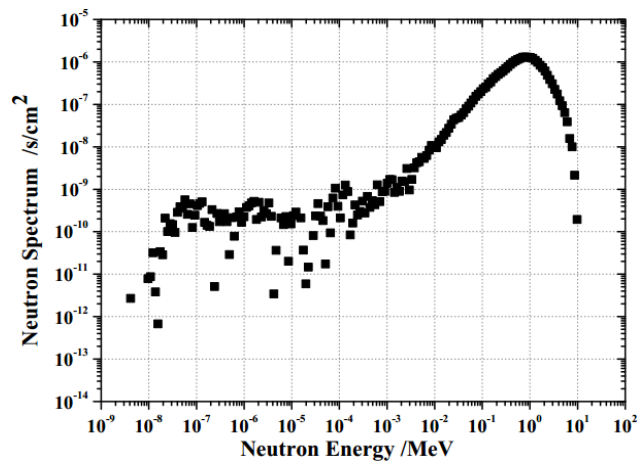

**Figure.6. Neutron energy spectrum in tungsten target for electron energy of 16 MeV**

In order to verify the design arrangement of neutron and γ rays guide tubes (as shown in Fig.2), the angular distribution of neutron and γ rays in the target housing are simulated. The point detectors are set around the target every $15^0$ to form a cycle, which are used to record the angular distribution information. The result is shown in Fig.7. It indicates that, the angular neutron flux distribution in the target housing is almost isotropic, and the γ rays flux in the direction of $0^0$ - $90^0$ and $270^0$ - $360^0$ is highest, following by $180^0$ direction. Considering the space limitation of experimental hall and experimental requirements, the best layout is to set the neutron guide tube perpendicular to the incident electron beam and the γ rays guide tube in forward direction of the incident electron beam, which further verifies the reasonable of the existing layout.

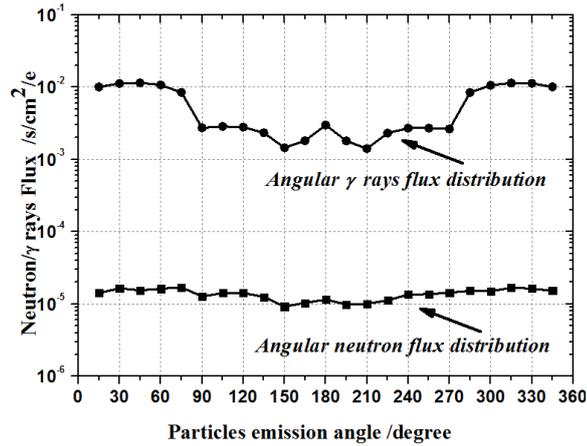

**Figure.7. Angular neutron/γ rays distribution of 0 - 360$^0$ in the plane of target chamber**

## 5. Conclusions

In this work, the subsection method combined with the variance reduction techniques based on Monte Carlo method which is called Sub-Var method is introduced to increase the efficiency and accuracy of the simulation. When the electron energy is 16 MeV and the power is 1050 W, the neutron flux and energy spectrum at TOF detector place are obtained and the result shows a good agreement with the experimental results. The calculation efficiency is increased by 23 times compared with the normal Monte Carlo method simulation. By this method, the thermal neutron detection efficiency of the $^6$LiF(ZnS) scintillator in PNS is calculated and the value is about 60.9% which is almost same with the calibration, the neutron flux at different place away from photoneutron target are also calculated that the available neutron flux for the integral experiment is up to $10^4$ /s/cm$^2$. It can be concluded that the Sub-Var method has a high capacity to modeling the PNS and calculate the neutron physical parameter, which is helpful to optimize the shielding scheme, verify the system design and provide more information for the future experiments.

### Acknowledgments

*This work is supported by the Chinese TMSR Strategic Pioneer Science and Technology Project under Grant No. XDA02010000, the National Natural Science Foundation of China under Grant No. 11475245, 91326201 and the Frontier Science Key Program of the Chinese Academy of Sciences under Grant No. QYZDY-SSW-JSC016.*